\documentclass{article}

\usepackage{arxiv}

\usepackage[utf8]{inputenc} 
\usepackage[T1]{fontenc}    
\usepackage{hyperref}       
\usepackage{url}            
\usepackage{booktabs}       
\usepackage{amsfonts}       
\usepackage{nicefrac}       
\usepackage{microtype}      
\usepackage{lipsum}
\usepackage{graphicx}
\title{Electrical Energy Consumption Control in Buildings Using Wireless Sensor Networks}

\author{
  Djamel Djenouri \\
  CERIST Research Center\\
  Algiers, Algeria.\\
  \texttt{ddjenouri@acm.org } \\
   \And
 Roufaida Laidi \\
  CERIST Research Center\\
  Algiers, Algeria.\\
  \texttt{rlaidi@cerist.dz} \\
  \AND
  Cherif Zizoua \\
  CERIST Research Center\\
  Algiers, Algeria.\\
}

\begin{document}
\maketitle

\begin{abstract}
Energy consumption in residential and commercial 
buildings has increased dramatically worldwide in the last decade, due 
to the constant population and economic growth, the proliferation of 
electronic and consumer appliances. This has dramatic footprint on the 
environment in terms of carbon emission, in addition to the economic 
impact. Green and smart building strategies will play a pivotal role to 
reduce this footprint and maximize economic and environmental 
performance. These strategies can be integrated into buildings at any 
stage, from design and construction, to maintenance and renovation. The 
use of modern Information and Communication Technologies (ICT), notably 
IoT solutions, for building control is one of the promising strategies 
for the future. The aim of this project was to explore this domain, and 
as a first step to develop a wireless sensor networks based solution for 
monitoring and energy management in offices. A prototype has been 
targeted as a proof of concept where sensors monitor physical parameters 
in CERSIT offices (presence of people, ambient light, etc.), and 
accordingly actuate lighting, air conditioning, etc. This report is a 
short summery of the different parts developed in this project.
\end{abstract}
\textbf{PROJECT IDENTIFICATION:}

\begin{table}[ht]
\centering
\begin{tabular}{|l|l|}
\hline
\textbf{Type} & Applied, development, Training \\
\hline
\textbf{Title} & \textbf{Electrical Energy Consumption Control in 
Buildings } \\
 & \textbf{ Using Wireless Sensor} \\
\hline
\textbf{Project Head} & Djamel Djenouri \\
\hline
\textbf{Starting} & Juliet 2014 \\
\hline
\textbf{End Date} & December 2017 \\
\hline
\textbf{Partner} & NTNU, Trondheim, Norway \\
\hline
\end{tabular}
\end{table}

\begin{table}[ht]
\centering
\begin{tabular}{|l|l|l|l|l|}
\hline
\textbf{Participants } \\
\hline
\textbf{Name} & \textbf{Title} & \textbf{Quality }(permanent, 
 & \textbf{Group} & 
\textbf{\% of } \\
&  &  adjunct, collaborator, &  & 
\textbf{participation} \\
& & extern student, etc.) & & \\
\hline
Djamel Djenouri  & DR & Permanent & WSN & 40\% \\
(Project PI) &  &  & & \\
\hline
Cherif Zizou & Ing. & Permanent & WSN & 60\%  \\
& & (absent for 18 months) & & during 1 year \\
\hline
Sahar Bouelkaboul & AR & Permanent & WSN& 40\%  \\
& & & & during 6 months\\
\hline
Roufaida Laidi & Ing & PhD Student  & WSN & 100\% 
\\
&  & (starting Jan 2016) & & 
during 20 months \\
\hline
Othmane Mokhtari &Ing & Engineer  & WSN & 100\% 
\\
&  &  & & 
during 5 months \\
\hline
\multicolumn{5}{|c|}{Alumni Group members Participating} \\
\hline
Messaoud Doudou & MR(B) & Permanent  & WSN & 40\% 
 \\
 &  & (absent for 2 years) & & 
during 1 year \\
\hline
Nouredine Lasla & MR(B) & Permanent  & WSN & 30\% 
\\
&  & (absent for 1 years) & & 
during 2 year \\
\hline
Abdelraouf Aoudjaout & AR & Permanent  & WSN & 
30\%  \\
& & (absent for 21 months)& & during 1 year\\
\hline
Miloud Baga & MR(B) & Permanent  & WSN & 40\%  \\
& &(absent for 2 years) & &during 
1 year \\
\hline
\end{tabular}
\end{table}

\textbf{RESULTS: }

A working prototype has been developed, featuring, optimal deployment 
for coverage, optimal duty cycling, realtime actuation for air 
conditional and lighting control. The prototype has been tested as 
proof-of-concept at the CEO office, and at another office (Building A) 
as well for harvesting long term data to evaluate the proposed 
solutions. Significant scientific results have been subject to 
publications in top tiered journals and conferences $[$1-6$]$, which 
will be presented in the following.

\textbf{Description of the prototype and related solutions }

The prototype mimics a pervasive system that can be integrated in 
existing buildings without any complicated wiring or setting (Fig.1). 
Realistic constraints are considered for this purpose such as 
sensing-hole, battery limitation, user comfort, daylight harvesting, 
etc. To ensure maximum coverage in presence of holes, the optimal 
placement of PIRs is formulated as a mixed integer linear programming 
optimization problem (MILP). Experimentations have been carried out to 
quantify the effects of the holes on the detection accuracy and to 
demonstrate the impact of the optimal PIRs placement on energy 
consumption. To facilitated installation and integration without 
complicated settings, notably in existing buildings, the system is 
designed to be battery operated. Therefore, energy efficiency will not 
be limited to optimize energy consumption in buildings, but also to 
optimize consumption in the components of the system (sensors and 
actuators). Duty cycling is inevitable to extend the network lifetime of 
such components, but the setting of this cycle yields a trade-off in 
optimizing the energy consumption i) at the building level, vs., ii) 
that consumed by sensors and actuators. Reducing energy consumption 
(duty cycle) of sensors/actuators will delay non-occupancy detections 
and thus will increase the building energy wastage, and vice-versa. Duty 
cycling the radios is dealt with and modeled as a cooperative game, 
which allows to derive a Nash Bargaining as the optimal balancing cycle. 
The proposed approach is analytically investigated using realistic 
parameters of the existing hardware and users' comfort. The results 
demonstrate that the system can survive for several years without battery replacement.

 \begin{figure}[ht]
 \centering
 \includegraphics[width=12.73cm,height=8.55cm]{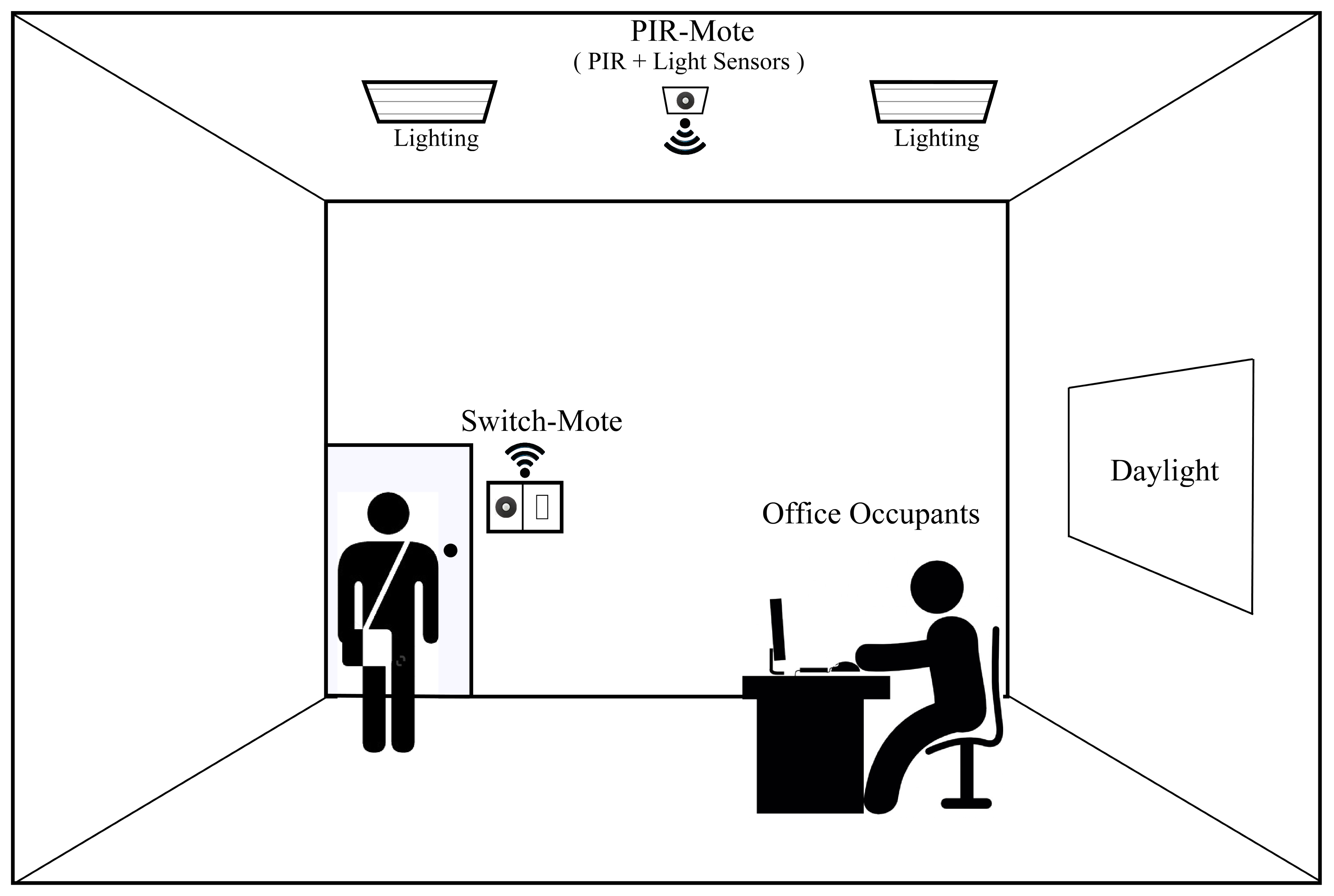}
 \caption{}
\end{figure}

The prototype includes:

\begin{itemize}
\item Sensor devices, including PIR, temperature, light, that are 
connected to actuators (smart switches).
\item A solution on PIR sensor deployment for optimal coverage in 
offices.
\item Optimal setting of parameters to enable effective usage of 
batteries, with lifetime estimated to up to 5 years for the sensor mote 
and 9 years for the actuator (smart switch).
\item Sensor and relay node indoor deployment by considering realistic 
physical layer parameters in building environment.
\end{itemize}
\textbf{Sensor nodes deployment}
PIRs are low-power sensors that use pyroelectric transducers, which 
convert infrared radiations into electrical signals. To increase the PIR 
sensitivity, a \textit{Fresnel lens} is used. It concentrates infrared 
radiations onto the detector. This results in a field-of-view (FoV) that 
is more like a discrete set of beams or cones with many sensing-holes. To be detected, the movements of the person should 
take place within the FoV. Fig.2(a) illustrates the different types of 
motion made by a human and the corresponding maximum sensing-hole size 
for which the motion can be detected by a PIR. The sensing-holes should 
not exceed 0.6 m to ensure an efficient detection of a sitting person's 
hand motions. The size and distribution of the holes impact the 
granularity of the PIR detections. Fig.2(b) illustrates the projection 
of the actual FoV of a Panasonic EKMB PIR sensor on a two-dimensional 
plane. The PIR is placed at the ceiling of an office and the projection 
is performed on the plane parallel to the ground and elevated at a 
typical height of desks, where most of persons' low movement activities 
take place (e.g. arm and hand movement when sitting). The figure shows 
the presence of several sensing-holes that represent more than 87\% of 
the total monitored office area, and their sizes vary from one region to 
another within the PIR's FoV. They may exceed 1m in some areas. These 
large sensing-holes may affect PIR-based occupancy detection systems and 
cause incorrect decisions, such as turning off a light or HVAC in the 
presence of a person, which limits the credibility of the system.

 \begin{figure}[ht]
 \centering
 \includegraphics[width=15.00cm,height=7.57cm]{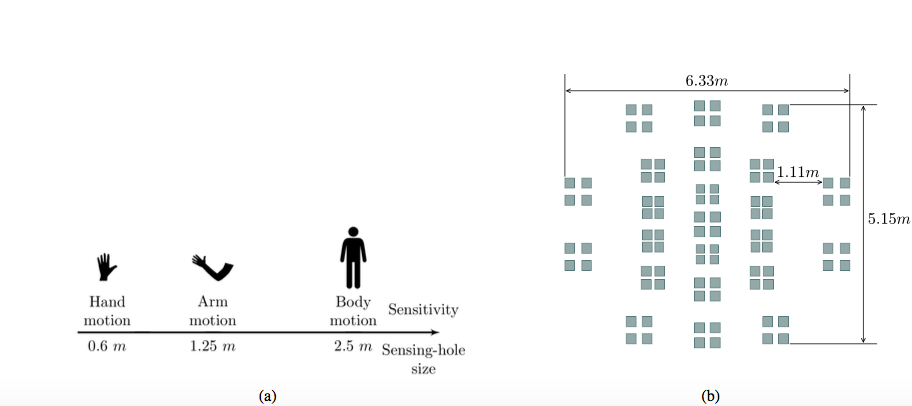}
\caption{}
\end{figure}

To deal with the sensing hole problem, we formulate a Maximum PIR 
Coverage (MPC) problem that finds the optimal positions of the PIRs for 
maximum coverage in the area of interest while considering the 
sensing-hole. To simplify the formulation, we consider the projection of 
the covered area on a two-dimensional plane as explained before. Despite 
such simplification, the computation of the detection zones for a given 
set of PIRs is difficult to formulate mathematically. The monitored area 
is discretized and considered as a set of points, where a point will be 
considered covered if it is within the coverage zone of at least one 
PIR. The obtained formulation is a mixed linear program to which we 
applied the Big-M method for transformation into standard form.

We have deployed an experimental PIR-based occupancy detection system to 
monitor an office and quantify the impact of the sensing-holes on the 
performances of the system. The PIR we used (EKMB PIR sensors from 
Panasonic) has been integrated to an nRF51-based mote (by Nordic 
Semi-conductors), which features a low-power SoC that embeds an ARM 
Cortex-M0 MCU, and a 2.4GHz wireless transceiver. We integrated the 
sensor to the mote via the available pins. The considered deployment 
area is a single-occupant office of 3.3 $\times $ 2.4 m$^{2}$. Most 
activities are concentrated over the office desk that received greater 
weights in the corresponding entries in the weighting matrix (denoted $\phi$). The discretization step was fixed to 0.3 m resulting in a grid of 11 
$\times $ 8 points. While the whole the office space falls within the 
sensing range, the real covered space is not the continuous space over 
this range but includes gaps (sensing- holes that has been explained 
above), and it might be represented as a set of discontinued squares 
(Fig.2.b). Three deployments scenarios have been evaluated. The first 
one corresponds to the optimal solution of the MPC problem when using 
one PIR (Fig.3). This deployment covers nearly 63\% of the desk's area. 
Optimal full coverage of this space is ensured with 3 PIRs, which 
corresponds to our second deployment scenario depicted in Fig.4. In the 
third scenario, a single PIR was placed using hole-unware placement as 
shown in Fig.5. It shows the real impact of sensing-holes on the 
performances of the detection system. It is worth noting that existing 
solutions consider the deployment represented by the third scenario as 
optimal since they ignore the presence of the sensing-holes and consider 
the PIR covers the whole space within its sensing range (all the 
office). This is effective in space with high motion (halls, doors, 
etc.), but not offices.

 \begin{figure}[ht]
 \centering
 \includegraphics[width=7.05cm,height=5.29cm]{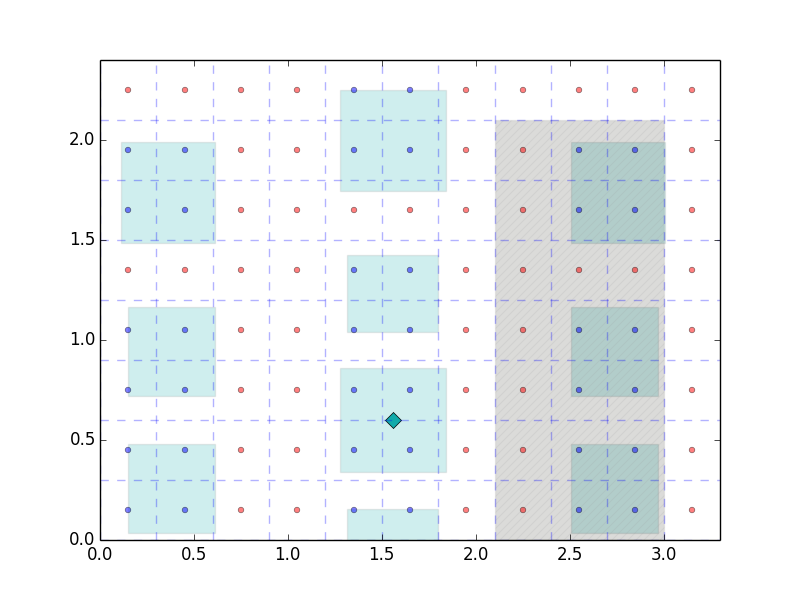}
 \caption{}
 \end{figure}
 \begin{figure}[ht]
 \centering
 \includegraphics[width=7.00cm,height=5.25cm]{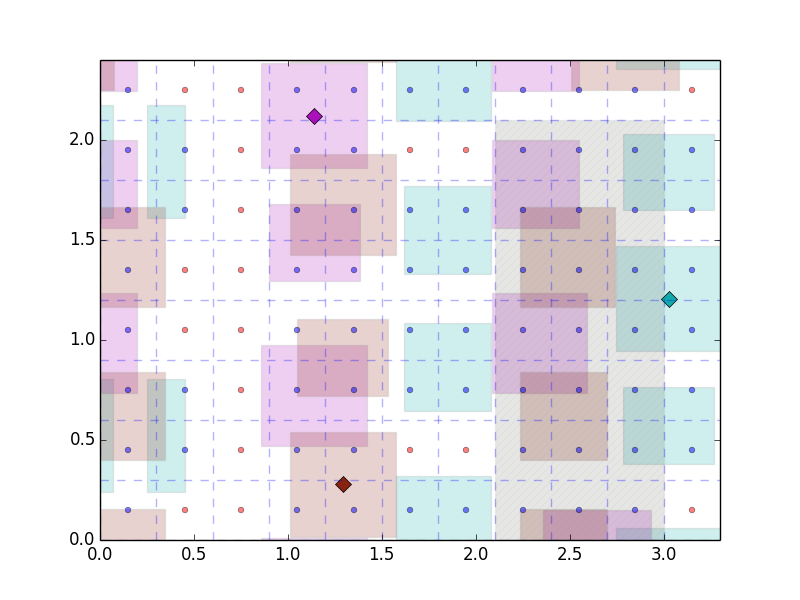}
 \caption{}
 \end{figure}
 \begin{figure}[ht]
 \centering
 \includegraphics[width=6.88cm,height=5.16cm]{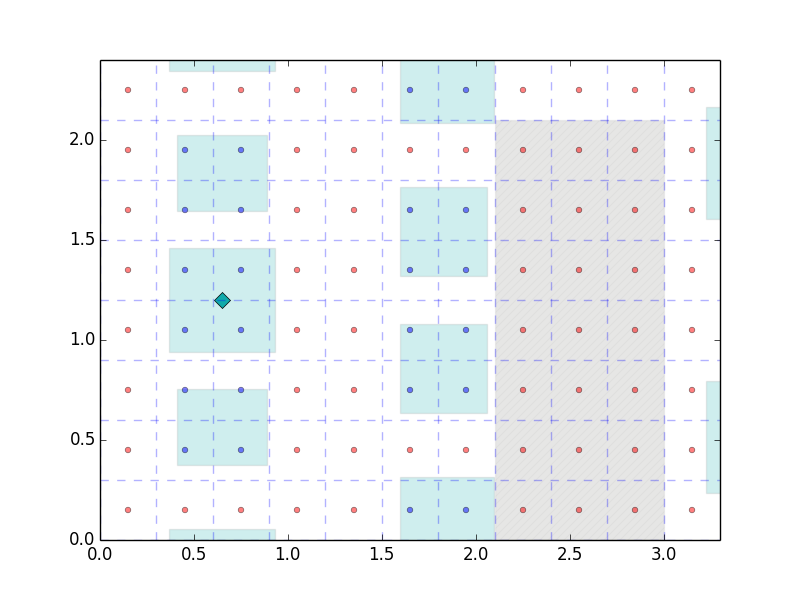} \caption{}
 \end{figure}

 To evaluate the performance of the system in the three deployment scenarios and under different time- out values (continuous time of absence reports by sensors before considering the space is vacant), we have measured two metrics, i) the comfort level, and, ii) the waste in energy usage. The first metric quantifies the ability of the system to preserve the convenience of users. That is, the ability not to disturb the occupants by keeping office energy supply on when they are present in the target area (i.e., ability to overcome false absence (FA). The second metric reflects the proportion of time the system fails to effectively detect (or react to) the absence of occupants, which implies a missed opportunity to reduce the energy consumption.
 Fig: 6 plots the Cumulative Distribution Function (CDF) of correct absence decisions as a function of the required time to take that decision (timeout). The results show fast convergence of the proposed solution vs. hole-unaware deployment. These results help in selecting the corresponding timeout to achieve a particular true absence (TA) probability (percentage). For instance, to realize 90\% of TA, the timeout should be set to 20sec, 35sec and 80sec for optimal3, optimal1, and hole-unaware deployment scenarios, respectively.
 
 \begin{figure}[ht]
 \centering
 \includegraphics[width=8.5cm,height=5.95cm]{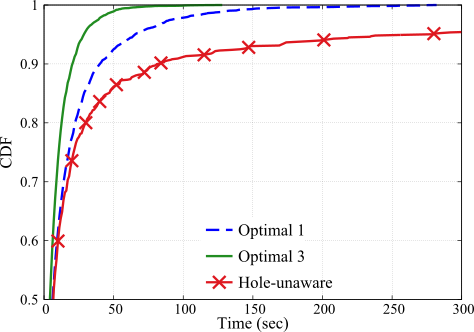}
 \caption{}
 \end{figure}
 
 We can also notice from Fig. 7 that the performance of the optimal solution using only one PIR is very close to the optimal full coverage solution using three PIRs. Compared to hole-unaware solution, Optimal3 allows to reduce energy wastage up to 9\% for high comfort level, and Optimal1 to up to 7.5
 
 \begin{figure}[ht]
 \centering
 \includegraphics[width=9.74cm,height=6.32cm]{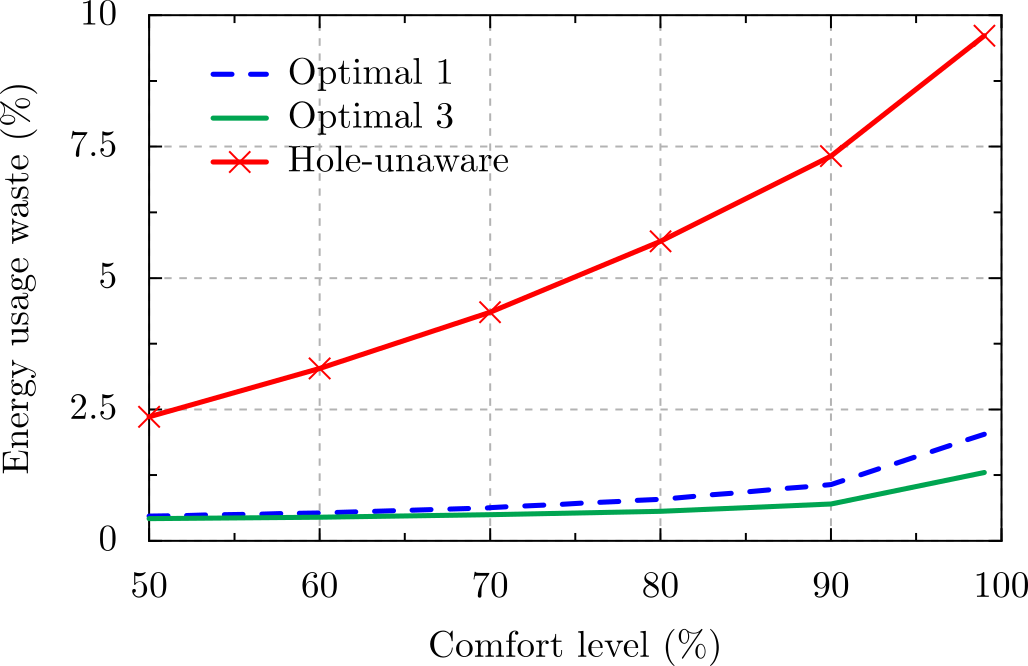}
 \caption{}
 \end{figure}
 We also considered placement of relay nodes (RNs) in the building environment and we proposed an original solution that consists of: i) the usage of a realistic physical layer model basedon a Rayleigh block- fading channel, ii) the calculation of the signal-to-interference- plus-noise ratio (SINR) considering the path loss, fast fading, and interference (which reflects the indoor environments), and iii) the usage of a weighted communication graph drawn based on outage probabilities determined from the calculated SINR for every communication link. Overall, the proposed solution aims for minimizing the outage probabilities when constructing the routing tree, by adding a minimum number of RNs that guarantee connectivity. In comparison to the state-of-the art solutions, the conducted simulations reveal that the proposed solution exhibits highly encouraging results at a reasonable cost in terms of the number of added RNs. The gain is proved high in terms of extending the network lifetime, reducing the end-to-end- delay, and increasing the goodput. While the prototype has only been deployed in single offices so far, the solution of optimal RNs placement will be useful when extending the prototype to the building scale, notably in multiple-floors buildings.

 \textbf{Extending system lifetime } 
 
 As the proposed control system relies on battery-operated sensor motes, it is important to optimize the battery power usage. Each mote should switch to power-save mode during inactivity periods. Since the radio is the most energy hungry component, the medium access control (MAC) protocol plays a key role in extending the system lifetime, by controlling the radio states and by employing low duty-cycles. In our occupancy detection scenario, the sensor-mote requires to report its new PIR or ambient light readings to (for example) the switch-mote to instantly turn on the light when the space becomes occupied or when the ambient light level becomes undesired. Instantaneous reporting is required to meet the expected users’ comfort. Because the moment when the occupancy state changes is unknown, the radio transceiver of the switch-mote should be always in standby (receive mode). This causes waste of an important amount of energy given that consumption in the reception mode is significant. The trivial solution to this problem is the implementation of a low duty-cycle MAC protocol, where energy saving is achieved by repeatedly switching the radio between active and sleep modes. This is known in literature by duty-cycling. In active mode, a node can receive and transmit packets, while in the sleep mode, it completely turns off its radio to save energy.  We targeted the use of existing duty-cycled MAC solutions (LPL used for illustration), while optimizing operation parameters. To ensure the users’ comfort required by the automatic light control system and extend its lifetime, we proposed to embed the switch-mote with a low cost PIR sensor. The later will be responsible of triggering the switch-mote once a movement is detected. By placing the switch-mote within the light-switch, next to the space entry, the new PIR sensor will be able to capture any entry and thus, enable the system to instantly turn on the light when the space becomes occupied. In this case, the switch-mote’s transceiver can be turned off without affecting the users’ comfort requirement. However, because the FoV of the new PIR sensor used by the switch-mote is mainly directed towards the space entrance, the latter cannot autonomously determine if the space is actually unoccupied (i.e., cannot rely on its PIR for that). This information can be only provided by the sensor-motes that have an appropriate coverage of the monitored space, i.e deployed using the solution described previously.
 
Both the sensor-mote and switch-mote are in power save mode with the radio turned off when the office is unoccupied. When a person enters the office, the new PIR sensor will trigger the switch-mote to immediately turn on the light and start duty-cycling the radio to receive occupancy state or ambient light reading from the sensor-mote. Whenever the later detects the activity in the office or read a new light value, it will activate its radio and start sending the new information to the switch-mote using the packetized preamble model similarly to the LPL scheme. The switch-mote keeps duty-cycling its radio until receiving an absence state. The sensor-mote returns to sleep mode after sending the new state. When an absence is detected, the sensor-mote reactivates its radio and reports the switch-mote in order to turn off the light and enable it go to sleep mode to save energy. The duty cycle period of the switch-mote has a conversely effect on the two motes’ lifetime. To calculate the optimal value of the wake-up period that enables making a balance between the sensor- mote lifetime and that of the switch-mote, we formulated the problem using game theory. We used the Bargaining model to define our two-player game. Instead of defining the individual nodes as players– which is common in the literature. The game players in our model are the systems objectives (sensor-mote and switch-mote lifetime). This limits the number of players and makes it independent from the problem size, which is scalable. The utility function of each player is used by the model to determine the optimal wakeup period parameter. Each player threats the other with using his best optimal point obtained from a non-cooperative game in which the player finds his best optimal operating value, i.e., player sensor-mote obtains its longest lifetime at the cost of decreasing the switch mote and vis versa. A bargaining game is then defined in order to find an agreement operational point that satisfies both players.  We considered a weighted model, where the desired sensor-mote lifetime is given a weight, i.e., it is $\alpha$ times more important than the switch-mote life- time ($\alpha \ge 1$). Fig.8 depicts the obtained results (lifetime and duty cycle parameter) for values of $\alpha$ ranging from $1$ to $5$. The figure shows that sensor-mote lifetime can reach more than 9 years without battery replacement for $2$ years lifetime for switch-mote, which is a tolerable frequency of replacing batteries of the switch that is usually more accessible than sensor-mote.  
 \begin{figure}[ht]
 \centering
 \includegraphics[width=13.83cm,height=9.26cm]{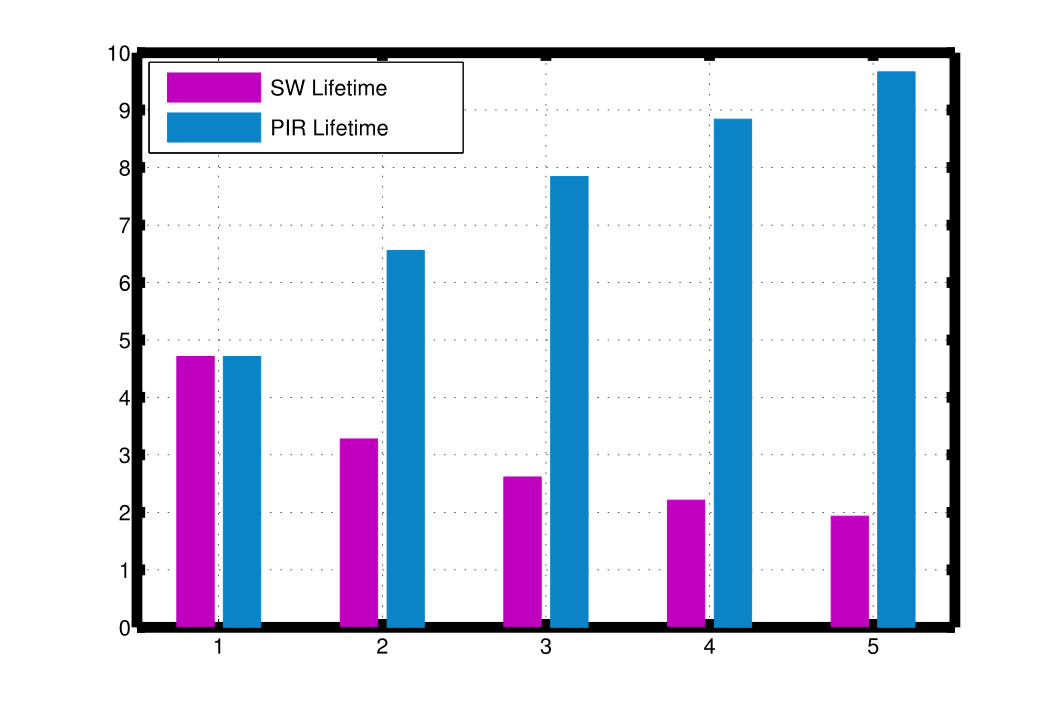} \caption{}
 \end{figure}
 \\
 \textbf{TRAINING: }
 \\
 \textbf{Undergraduate internships: } One student for short internships (1 month).
 \\
 \textbf{Masters: } Three students for short internships (1 month).
 \\
 \textbf{ Doctorate: } Two students working on the projet (ongoing).
 \\
 \textbf{ COLLABORATION: } NTNU (Norwegian University of Science and Technology), Trondheim, Norway. Mobility Grant has been obtained from The Norwegian Research Council to cover research visits to NTNU.

 \textbf{ DIFFICULTIES MET: } Completely Unstable Human Resources, with too many long leaves (for a year and more) of staff simultaneously. Several risk management measures have been taken to deal with this problem. Many researchers have been involved midterm during the project lifetime. Further, some objectives have been cut down (generalization of the prototype to several buildings).

 \textbf{PERSPECTIVE: } The project has been limited to sensing and realtime actuation for energy control in offices. Considering jointly energy optimization and user comfort services is an interesting perspective. Here by comfort, we are not taking about optimal timeout values to avoid disturbing users with false actuations (which has already been dealt with in the project), but we refer to more advanced services such as including security services, customized AC, etc. Moreover, sensing has been limited to the use of sensor motes. Integrating sensing through other IoT devices (smart phones), smart meters, information from the cloud etc., will be required to achieve the above mentioned advanced services.

 \textbf{REFERENCES }
 
 \begin{enumerate}
 \item M. Bagaa, A. Chelli, D. Djenouri, T. Taleb, I. Balasingham, K. Kansaneng, "Optimal Placement of Relay Nodes Over Limited Positions in Wireless Sensor Networks", EEE Transactions on Wireless Communications 16(4): 2205-2219, 2017.
 \item D.Djenouri, M. Bagaa, "Synchronization Protocols and ImplementationIssues in Wireless Sensor Networks: A Review", IEEE Systems Journal, 10(2), pp 617-627, June 2016. 
 \item A. Chelli, M. Bagaa, D. Djenouri, I. Balasingham, T. Taleb, "One-Step Approach for Two-Tiered Constrained Relay Node Placement in Wireless Sensor Networks", IEEE Wireless Communications Letters, Vol 5(4), pp 448-451, August 2016. 
 \item M. Doudou, J. M. Barceló-Ordinas, D. Djenouri, J. G. Vidal, A. Bouabdallah, Nadjib Badache: Game Theory Framework for MAC Parameter Optimization in Energy-Delay Constrained Sensor Networks. ACM Trans on Sensor Networks 12(2): 10, May 2016.
 \item Abdelraouf Ouadjaout, Noureddine Lasla, Djamel Djenouri, Cherif Zizoua: On the Effect of Sensing-holes in PIR-based Occupancy Detection Systems. SENSORNETS, Rome Feb, PP 175-180, Feb 2016.
 \item D. Djenouri, M. Bagaa: Implementation of high precision synchronization protocols in wireless sensor networks. IEEE WOCC 2014.
 \item D. Djenouri: MLE for Receiver-to-Receiver Time Synchronization in Wireless Networks with Exponential Distributed Delays. IEEE VTC Spring 2014: 1-5.
 \item	R. Laidi, D. Djenouri, "UDEPLOY: User-Driven Learning for Occupancy Sensors DEPLOYment In Smart Buildings", IEEE PerCom Workshops, March 2018 (accepted). 

 \end{enumerate}

\end{document}